\documentclass[dvips, twocolumn, prb, superscriptaddress, showpacs, floatfix, amsmath, amssymb]{revtex4}
\usepackage{graphicx}

\begin{document}
\title{Optically induced coherent intra--band dynamics in disordered
semiconductors}
\author{C. Schlichenmaier}
\affiliation{Fachbereich Physik und Wissenschaftliches Zentrum f\"ur
	Materialwissenschaften, Philipps--Universit\"at, 
  Renthof 5, D--35032 Marburg, Germany}
\author{I. Varga}
\affiliation{Fachbereich Physik und Wissenschaftliches Zentrum f\"ur 
	Materialwissenschaften, Philipps--Universit\"at,  
  Renthof 5, D--35032 Marburg, Germany}
\affiliation{Elm\'eleti Fizika Tansz\'ek, Fizikai Int\'ezet, 
  Budapesti M\H uszaki \'es Gazdas\'agtudom\'anyi Egyetem, 
  H--1521 Budapest, Hungary}
\author{T. Meier} 
\affiliation{Fachbereich Physik und Wissenschaftliches Zentrum f\"ur 
	Materialwissenschaften, Philipps--Universit\"at,  
  Renthof 5, D--35032 Marburg, Germany}
\author{P. Thomas} 
\affiliation{Fachbereich Physik und Wissenschaftliches Zentrum f\"ur 
	Materialwissenschaften, Philipps--Universit\"at,  
  Renthof 5, D--35032 Marburg, Germany}
\author{S. W. Koch}
\affiliation{Fachbereich Physik und Wissenschaftliches Zentrum f\"ur 
	Materialwissenschaften, Philipps--Universit\"at,  
  Renthof 5, D--35032 Marburg, Germany}
\date{\today}
\begin{abstract}
On the basis of a tight--binding model for a strongly disordered semiconductor with
correlated conduction- and valence band disorder a new coherent dynamical intra--band effect is analyzed. For systems that are excited by two, specially designed ultrashort light--pulse sequences delayed by $\tau$ relatively to each other echo--like phenomena are predicted to occur. In addition to the inter--band photon echo which shows up at exactly $t=2\tau$ relative to the first pulse, the system responds with two spontaneous intra--band current pulses preceding and following the appearance of the photon echo. The temporal splitting depends on the electron--hole mass ratio. Calculating the population relaxation rate due to Coulomb scattering, it is concluded that the predicted new dynamical effect should be experimentally observable in an interacting and strongly disordered system, such as the Quantum--Coulomb--Glass. 
\end{abstract}
\pacs{72.40.+w; 78.47.+p; 72.80.Ng}
\maketitle

\section{Introduction}
Echo phenomena, the prototype being the Hahn spin echo \cite{hahn} for spin--1/2--systems, rely on the generation of a coherent ensemble of excitations with a continuous distribution of frequencies. After pulsed excitation the macroscopic response of the ensemble decays as a consequence of destructive interference effects in the continuum of excited frequencies. A second, delayed excitation pulse induces a rephasing of the individual excited species such that at twice the delay time the ensemble shows a spontaneous macroscopic response, the echo. A necessary requirement for the observability of this coherent dynamics are sufficiently weak dephasing interactions on the time scale of the pulse delay. The microscopic reason for the appearance of an echo is that the second pulse causes phase conjugation of the coherent excitation generated by the first pulse. In close analogy to the spin echo also photon echoes have been observed in ensembles of two--level absorbers \cite{hartmann}. There are also phonon echoes \cite{phonon} and temperature echoes \cite{temp}, even classical mechanic ensembles of pendulums can show echo phenomena \cite{pendel}. 

Photon echoes have also been studied in more complicated systems, such as ordered \cite{koch} and disordered \cite{goebel} semiconductors. The optically excited inter--band transitions in a semiconductor cannot be considered as an ensemble of independent two--level absorbers due to the strong interaction of the electron--hole pairs. Photon echoes therefore may show a decay as a function of the delay time due to the Coulomb interactions, due to disorder, and due to combined interaction--disorder effects \cite{koch, lonsky, bennhardt, weiser}. In Ref. \cite{nigge} a new echo phenomenon has been proposed for disordered conductors or Anderson insulators. On the basis of a non--interacting one--dimensional tight--binding band with diagonal disorder filled with a low density of carriers the current response to short externally applied voltage pulses was calculated. Assuming excitation with two short voltage pulses at $t = 0$ and $t=\tau$, where $\tau$ was chosen larger than the typical elastic scattering time $\tau_{el}$, it was predicted that the system spontaneously responds with a current pulse exactly at time $t=2\tau$. In contrast to spin echoes and photon echoes this current echo is not related to phase conjugation. 

The influence of Coulomb interactions on the current echo was investigated in Ref. \cite{zim}. The numerically exact calculation for a small tight--binding system showed that the current echo should remain visible in the presence of the many--body interaction, however so far no experimental observation has been reported. 

In a series of papers van Driel, Sipe, and coworkers have shown that in semiconductors currents can be induced optically on an ultra--short time scale using a coherent control scheme \cite{toronto}. In addition, a resulting current pulse could possibly be observed using THz--detection techniques. 

Stimulated by these results, we studied optically induced current phenomena in disordered model systems \cite{we}. By solving the equation of motion for the intra--band current in a noninteracting tight--binding model we concluded that on the basis of the coherent control scheme it should be indeed possible to generate a current pulse also in a strongly disordered semiconductor. The current pulse decays due to elastic scattering. In addition, the current traces show signatures of Anderson localization (in a one--dimensional noninteracting disordered system all states are localized). The application of two delayed optical pulses, both generating a current pulse, was found to result in a sizable echo in the intra--band current, that appears at $t=2\tau$. At the same time the inter--band polarization shows the conventional photon echo. 

Continuing our earlier work \cite{we} and making our model more realistic we were surprised to discover that the optically induced intra--band dynamics in a disordered semiconductor with correlated conduction- and valence--band disorder shows features that differ profoundly from both the photon and the current echo. By allowing the effective masses of the electrons and holes to differ from each other, we find that the related spontaneous signals appear at different times. While the inter--band photon echo always shows up at $t=2\tau$, the intra--band signals split into an advanced and a retarded signal current pulse. This splitting depends on the mass ratio. For the semiconductor model we find that these two current pulses appear only if the disorder is correlated for electrons and holes, which is often at least approximately valid in real low--dimensional semiconductor nano--systems where disorder may be due to the fluctuations of the confinement potential. 

We furthermore find that only the first excitation pulse at $t=0$ has to be designed according to the coherent control scheme, which is necessary to generate an intra--band current. For the second pulse at $t=\tau$ a normal resonant optical excitation is sufficient. This simple excitation pulse does not generate a current pulse at $t=\tau$, however, it is still able to initiate the pair of spontaneous signals. These signals are most prominent for high excitation density. Since the many--particle Coulomb interaction has been ignored in these calculations, one might suspect that Coulomb scattering leads to sufficiently rapid dephasing such that the new dynamics is no longer observable. The consideration of the Coulomb interaction for the situation at hand requires to calculate the response at least up to 5th--order in the external light field and to treat the interaction consistently with all the relevant correlations (for a 3rd--order treatment in a disordered semiconductor model see, e.g., \cite{sieh, weiser, dirk}). Even for a one--dimensional system this task is beyond present computer capacity. In order to get a feeling for the relevant time scales we consider a much simpler case. For a strongly disordered one--dimensional single--band model we calculate the population relaxation time due to Coulomb scattering by treating the many--particle interaction in second Born approximation on the basis of Hartree--Fock states. The situation envisaged is the so--called Quantum--Coulomb--Glass model, one of the most challenging problems in modern many--particle physics. The results we obtain can be taken as an indication that in fact strong disorder leads to a slowing--down of the dynamical Coulomb effects. For the time being this suggests that the current echo should be observable in experiment. 

This paper is organized as follows. In section \ref{model} the model is introduced. The relevant observables, the inter-- and intra--band currents and the polarization, are defined in section \ref{obs}. The equations of motion are given in section \ref{eqm} and the coherent control excitation scheme is described in section \ref{coco}. Finally, in section \ref{res} the numerical results are presented and discussed in section \ref{dis}. In the concluding section \ref{con} we consider the possible influence of the Coulomb many--particle interaction and give evidence for the observability of the predicted signals in experiments. 

\section{The model}
\label{model}
We consider a one--dimensional two--band tight--binding model with nearest neighbor coupling $J^{\lambda}$ defined to model a direct gap semiconductor. The $N$ sites $i$ have nearest neighbor separation $|\vec{R}|$ and periodic boundary conditions are applied. Every site $i$ at position $\vec {R}_i$ carries two energies $\epsilon_i^c$ and $\epsilon_i^v$. They are distributed randomly according to a box--shaped distribution of width $W^{\lambda}$. In the electron--hole picture all electron energies and hole energies $\epsilon_i^{\lambda}$ as well as the couplings $J^{\lambda}$ are positive quantities if the energy is taken to be zero in the gap between the valence and the conduction band. The Hamiltonian matrix for band $\lambda$ is then
\begin{equation}
T_{ij}^{\lambda}=\delta_{ij} \epsilon_i^\lambda-
J^\lambda\left(\delta_{ij-1}+\delta_{ij+1}\right). 
\label{matrix}
\end{equation}
For an ordered situation ($\epsilon_i^{\lambda}=\epsilon_0^{\lambda},\;W^\lambda=0\mbox{eV}$) we have a direct semiconductor with cosine bands and a gap in the center ($\vec k=0$) of the Brillouin zone. Close to $\vec k=0$ the bands are characterized by effective masses $m_{\lambda}$ related to the couplings $J^{\lambda}$
\begin{equation}
m_{\lambda}=\frac{\hbar^2}{2J^{\lambda}|\vec{R}|^2}.
\label{mass}
\end{equation}
In the disordered case the relevant disorder parameter is
\begin{equation}
\eta^\lambda = \frac{W^\lambda}{J^\lambda}. 
\label{defeta}
\end{equation}
In our one--dimensional system all single--particle states are localized for non--zero $W$. The disorder is called uncorrelated, if the site energies are distributed independently from each other in the two bands. It is called correlated if
\begin{equation}
\frac{\epsilon_i^c-\langle\epsilon_i^c\rangle}{J^c}=
\frac{\epsilon_i^v-\langle\epsilon_i^v\rangle}{J^v}. 
\label{corr}
\end{equation}
where $\langle\epsilon_i^{c, v}\rangle$ are the expectation values of the site energies. In this work we exclusively treat correlated disorder. It models, e.g., a disorder potential in a semiconductor heterostructure with effective dimensionality less than three, which is produced by local fluctuations in the confining potential. 

The total Hamiltonian then reads
\begin{eqnarray}
H &=& H_{0} + H_{I}, \label{ham}\\
H_{0} &=& \sum _{i, j\atop\lambda=c, v} \tilde T_{ij}^{\lambda}
a^{+}_{\lambda i}a_{\lambda j}, \\
H_{I} &=& -\vec{E}(t) \cdot \vec d. 
\end{eqnarray}
$a^{+}_{\lambda i}$, $a_{\lambda j}$ are electron creation an annihilation operators, respectively. The couplings $\tilde T$ in the electron picture are unchanged for the conduction band, i.e. $\tilde T_{ij}^{c} = T_{ij}^{c}$, whereas a change of sign is required for the valence band, i.e. $\tilde T_{ij}^{v} = -T_{ij}^{v}$. 
\begin{equation}
\vec {d}=\sum_{i\atop\lambda, \lambda'=c, v}
  \vec {d}_{ii}^{\lambda\lambda'}a^{+}_{\lambda i}a_{\lambda'\,i}
\label{pp}
\end{equation}
and
\begin{equation}
\vec {d}_{ii}^{\lambda\lambda'} = -e
 \left(\vec R_i\delta_{\lambda\lambda'} + \vec r_{\lambda\lambda'}\right) \label{ddd} 
\end{equation}
$\vec r_{\lambda \lambda'}$ is the inter--band optical dipole matrix element. 

\begin{table}
\caption{\label{tab1} Parameters of the semiconductor model and the light pulse}
\begin{ruledtabular}
\begin{tabular}{|l|c|}
$\langle\epsilon^c\rangle+\langle\epsilon^v\rangle$&1.316 eV	\\
$J^c$					&	34 meV		\\
$r_{cv}$				&	3.0 \AA		\\
$|\vec{R}|$				&	20 \AA		\\
Disorder $\eta$				&	2		\\
Number of sites				&	71		\\
\hline
Light field described in section \ref{coco}&			\\
$\phi_{12}$				&	$\pi/2$		\\
central wavelength of full--gap--pulse	&	1.316 eV
\end{tabular}
\end{ruledtabular}
\end{table}
\section{The observables}
\label{obs}

The total polarization is given by
\begin{equation}
\vec P= \frac{\langle \vec d\rangle}{V}
\end{equation}
and the total current by
\begin{eqnarray*}
\vec J &=& \langle\dot{\vec d}\rangle
\label{jj}\nonumber\\
&=& \frac{1}{i\hbar }\big \langle[{\vec d}, H_{0}]\rangle\nonumber\\
&=& \frac{ie}{\hbar}\sum_{\lambda\; ij}
\left(\vec{R}_i - \vec{R}_j\right) \tilde T^{\lambda}_{ij}
\langle a^{+}_{\lambda i}a_{\lambda\,j}\rangle\nonumber\\
&& {}+ \frac{ie}{\hbar}\sum_{\lambda\lambda'\; ij}
\vec r_{\lambda\lambda'}\left(\tilde T^{\lambda'}_{ij}-
\tilde T^{\lambda}_{ij}\right)\langle a^{+}_{\lambda i}
a_{\lambda'\,j}\rangle\nonumber\\
&=& \frac{ie}{\hbar}\vec R\sum_{\lambda i}\tilde J^{\lambda}\left(\langle
a^{+}_{\lambda i}a_{\lambda i+1}\rangle - \langle a^{+}_{\lambda i}
a_{\lambda i-1}\rangle\right) \nonumber\\
&& {}- \frac{ie}{\hbar}\sum_{\lambda\lambda'\; i}
\vec r_{\lambda\lambda'}\left(\tilde\epsilon^{\lambda}_{i}-
\tilde\epsilon^{\lambda'}_{i}\right)\langle a^{+}_{\lambda i}
a_{\lambda' i}\rangle \nonumber\\
&& {}+ \frac{ie}{\hbar}\sum_{\lambda\lambda'\; i}
\vec r_{\lambda\lambda'}\left(\tilde J^{\lambda} - \tilde J^{\lambda'}
\right) \\
&& \phantom{{}+ \frac{ie}{\hbar}\sum_{\lambda\lambda'\; i}
\vec r_{\lambda\lambda'}}\times\left(\langle a^{+}_{\lambda i}a_{\lambda'\, i+1}\rangle +
\langle a^{+}_{\lambda i}a_{\lambda'\, i-1}\rangle\right), 
\end{eqnarray*}
where $\vec R= \vec{R}_{i+1}-\vec{R}_i$. These observables have both intra- and inter--band contributions due to the first and second term in Eqn.~(\ref{ddd}), respectively. 

Denoting the expectation values of the operators by
\begin{subequations}
\label{el-loch}
\begin{eqnarray}
p_{ij} & = & \langle a^{+}_{vi}a_{cj}\rangle, \label{aband}\\
n^{e}_{ij} & = & \langle a^{+}_{ci}a_{cj}\rangle, \label{eqelect} \\
n^{h}_{ij} & = & \delta _{ij}-\langle a^{+}_{vj}a_{vi}\rangle. 
\label{eqloch}
\end{eqnarray}
\end{subequations}
where $p_{ij}$ is the inter--band coherence related to the inter--band polarization (and the inter--band current) and $n^{e}_{ij}$ and $n^{h}_{ij}$ are the intra--band coherences ($i \neq j$) and densities ($i = j$) related to the intra--band current. 

The intra--band current $\vec {J}^{intra}$, the inter--band current $\vec {J}^{inter}$, and the total polarization $\vec P$ are given by
\begin{eqnarray}
\vec{J}^{intra} &=& \frac{2e\vec R}{\hbar}\left[J^c\sum_{i}
\mbox{Im}\left[n^e_{i+1, i}\right] \right. \nonumber\\
& & \left. \phantom{\frac{2e\vec R}{\hbar}[J^c} {} - J^v\sum_{i}\mbox{Im}\left[n^h_{i+1, i}
\right]\right], \label{intra}\\
\vec{J}^{inter} &=& {}-\frac{2e\vec r_{cv}}{\hbar} \sum_{i}
\left(\epsilon^{e}_{i}+\epsilon^{v}_{i}\right)\mbox{Im}
\left[p_{ii}\right]\label{inter}\\
&& {}+ \frac{2e\vec r_{cv}}{\hbar} \left(J^{c}+J^{v}\right)
\sum_{i}\mbox{Im}\left[p_{i-1, i}+p_{i+1, i}\right], 
\nonumber\\
\vec P &=& -\frac{e}{V}\sum_{i}\Big[\vec R_{i}
\left(n^{c}_{ii}-n^{h}_{ii}+1\right) \nonumber\\
& & \qquad \qquad \quad {} + 2\vec r_{cv}\mbox{Re}\left[p_{ii}\right]
\Big], 
\label{pol}
\end{eqnarray}
respectively. In Eqn.~(\ref{pol}) the first and second terms refer to the intra-- and inter--band polarization, respectively. 

\section{Equations of motion}
\label{eqm}
Using the Heisenberg equation of motion and taking the expectation values everywhere we obtain
\begin{eqnarray}
\frac{d}{dt}p_{ij}
&=&
-\frac{i}{\hbar }\big (\epsilon^{v}_{i}+\epsilon^{c}_{j}\big )p_{ij} \label{bew}
\\
&+&
\frac{i}{\hbar}J^v\left(p_{i-1j}+p_{i+1j}\right) +
 \frac{i}{\hbar} J^{c}\left(p_{ij-1}+p_{ij+1}\right)\nonumber\\
&+&
\frac{i}{\hbar}e \vec{E}(t) \left[\left(\vec R_i-\vec R_j\right)p_{ij}
+ \vec r_{cv} \left(n^e_{ij} + n^h_{ji} - \delta_{ij}\right)\right], 
\nonumber \\
\frac{d}{dt}n^e_{ij}
&=&
\frac{i}{\hbar }\big (\epsilon ^{c }_{i}-\epsilon ^{c}_{j}\big )n^c_{ij} \label{bewe}
\\
&+&
\frac{i}{\hbar}J^c\left(n^e_{ij-1} + n^e_{ij+1} - n^e_{i-1j} - n^c_{i+1j}\right)\nonumber\\
&+&
\frac{i}{\hbar }e\vec E(t)\left[(\vec R_i-\vec R_j)n^c_{ij} +
\vec r_{cv}\left(p_{ij} - p^*_{ji}\right)\right], \nonumber \\
\frac{d}{dt}n^h_{ij}
&=& \frac{i}{\hbar}\big (\epsilon^{v}_{i}-\epsilon^{v}_{j}\big )n^h_{ij}\label{beweg}
\\
&+&
\frac{i}{\hbar}J^v\left( n^h_{ij-1} + n^h_{ij+1} - n^h_{i-1j} - n^h_{i+1j}\right)\nonumber\\
&-&
\frac{i}{\hbar }e\vec E(t)\left[\left(\vec R_i-\vec R_j\right)n^h_{ij}
+ \vec r_{cv}\left(p^*_{ij}-p_{ji}\right)\right] , \nonumber
\end{eqnarray}
which differ from the conventional optical Bloch equations for a noninteracting tight--binding model \cite{sieh, weiser, dirk} by the terms containing the position $\vec {R}_i$ of the sites. 

We numerically solve these equations using the standard 4th--order--Runge--Kutta algorithm. 

\section{Current excitation by coherent control}
\label{coco}
\begin{figure}
\includegraphics[width=3in]{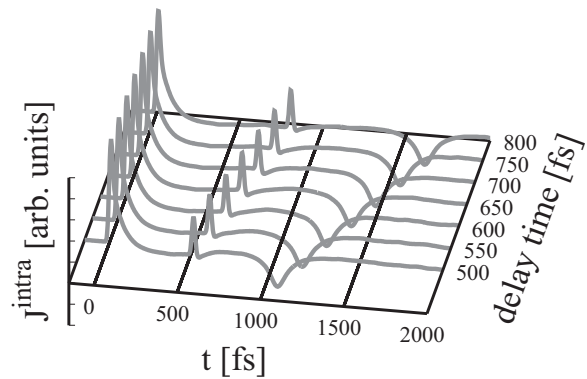}
\caption{\label{fig1} The negative peaks are spontaneous responses following a succession of two positive current pulses generated by coherent control. The effective masses of electrons and holes are identical. Note the correct shift of the spontaneous signal with $2\tau$. The exciting light pulses have the pulse area $A_n=0.64\pi$ (section \ref{excit}) and a duration of $t_{Ln}=20$fs. Each curve has been averaged over 150 disordered realizations. From Ref. \cite{we}. }
\end{figure}

We apply the coherent control scheme developed by the Toronto Group \cite{toronto} in order to optically generate a short intra--band current pulse in both bands. The first excitation at time $t=0$ is chosen to be due to a light field
\begin{eqnarray}
\vec E(t) &=& \vec E_1e^{-\left(\frac{t}{t_{L1}}\right)^2}
      \cos\left[\frac{\omega}{2}t\right]\nonumber\\ 
   &+& \vec E_2e^{-\left(\frac{t}{t_{L2}}\right)^2}
       \cos\left[\omega t + \phi_{1, 2}\right], 
\label{puls1}
\end{eqnarray}
where $\vec{E}_n$ are the amplitudes, $t_{Ln}$ the temporal widths and $\phi_{1, 2}$ is the relative phase of the two contributions having frequency $\omega$ (called full--gap pulse) and $\omega/2$ (called half--gap pulse). We take $\omega$ to be larger than the band gap, whereas $\omega/2$ is smaller than the band gap. In a previous paper \cite{we} we have shown, using the above model, that this excitation results in a current depending on $\sin[\phi_{1, 2}]$ that decays due to scattering at the disorder. This decay is modulated by oscillations which are a fingerprint of Anderson localization. It was also shown there (see Fig.~\ref{fig1}) that a second identical pulse, arriving at time $t=\tau$, leads to a spontaneous current echo at time $t=2\tau$ having opposite direction. In that work identical electron and hole masses have been taken, i.e. $J^c=J^v$. The process of generating a current is of at least third order in the field amplitudes. The current echo requires additional excitation with delayed pulses and is therefore of higher order. The analysis presented in the following section shows that the appearance of an echo is at least of fifth--order in the external fields. Although the current pulses have been generated optically, the dynamics seems to strongly resemble the current echo as it was originally suggested \cite{nigge} for a single band model with current generation by voltage pulses. As will be shown in the next section, however, the intra--band dynamics initiated in a semiconductor is profoundly different from the current echo in a single band situation. 

\section{Numerical results}
\label{res}
\subsection{Dependence on disorder and on electron--hole mass ratio}
There are three main findings which point out that the dynamics initiated by the coherent control excitation characterizes a new coherent phenomenon. 

($i$) A spontaneous signal pulse is completely absent if instead of correlated disorder according to Eqn.~(\ref{corr}) we consider a model with uncorrelated disorder. It should be mentioned, that in this case for a single disorder realization current fluctuations are excited even by normal band--band excitation. These can be suppressed only by extensive configurational averaging over a large number of disorder realizations. After sufficient averaging no detectable signal can be seen in the resulting current traces. 
\begin{figure}
\includegraphics[width=3in]{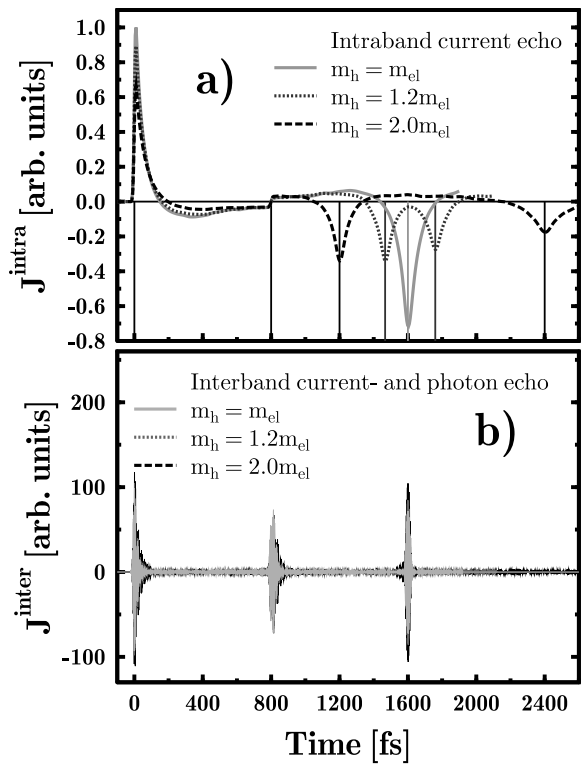}
\caption{{\bf a} Spontaneous response in the intra--band current after excitation at $t=0$ with coherent control pulses and at $t=800$fs with a single full--gap pulse. For different electron and hole masses the spontaneous response splits into a pair of signals according to Eqns.~(\protect\ref{eq_tv}), (\protect\ref{eq_tc}). {\bf b} The inter--band current (and also the polarization) for the same excitation scenario continue to rephase at $t=2\tau$. Pulse areas are $A_1=0.5\pi$, $A_2=0.5\pi$, $A_4=\pi$, the pulse duration $t_{Ln}=10$fs. The curves are averages over 64 disordered model realizations. }
\label{fig2}
\end{figure}

($ii$) The spontaneous response results even for a simplified excitation sequence. The first pulse at $t=0$ is taken to generate a current, i.e. it is given by Eqn.~(\ref{puls1}). However, for the second pulse at $t=\tau$ we take a simple full--gap pulse with only a single central frequency $\omega$. In order to have an influence on the dynamics of intra--band quantities, this pulse has to enter the equations of motion at least in second order. As a result we still see a spontaneous response at the time $t=2\tau$, see Fig.~\ref{fig2}a. We conclude, therefore, that the spontaneous signal does not require excitation by two current pulses. Instead only one current pulse is necessary and the second light pulse can be a single full--gap pulse. The current pulse has to precede the full--gap pulse, since the reverse order of pulses does not yield a spontaneous response. Because of this finding we avoid to call the spontaneous response an echo in the following. 

($iii$) For different electron and hole masses Eqn.~(\ref{mass}), $J^v/J^c < 1$, we find that, surprisingly, there are two separate intra--band responses, one preceding $t=2\tau$ and one following $t=2\tau$. In particular, the delayed contribution is due to the valence band and appears at
\begin{equation}
t_v=\left (1+\frac{J^c}{J^v}\right )\tau
\label{eq_tv}
\end{equation}
while the preceding contribution is due to the conduction band and appears at
\begin{equation}
t_c=\left (1+\frac{J^v}{J^c}\right )\tau. 
\label{eq_tc}
\end{equation}

Note that the inter--band photon echo in all these cases always appears at $t=2\tau$, as shown in Fig.~\ref{fig2}b. 

\subsection{Dependence on excitation density}
\label{excit}
\begin{figure}
\includegraphics[width=3in]{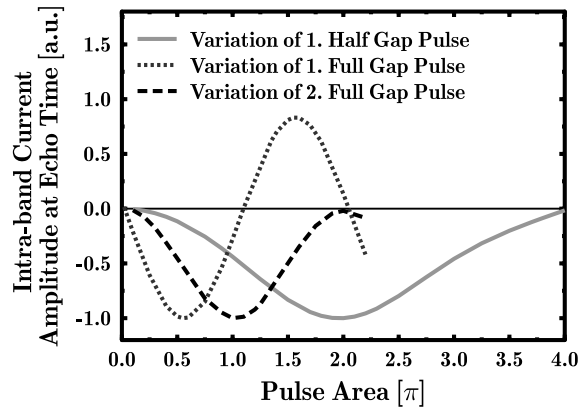}
\caption{Dependence of the amplitude of the spontaneous signal on the pulse areas. Solid line:
variation of 1. half--gap--pulse, $A_2=0.1\pi$, $A_4=0.2\pi$; dashed line:
variation of 1. full--gap--pulse, $A_1=0.1\pi$, $A_4=0.2\pi$; dotted line:
variation of 2. full--gap--pulse, $A_1=0.1\pi$, $A_2=0.1\pi$. $t_{Ln}=10fs$
in all cases. The data has been extracted from averages over 64
disorder realizations. The effective masses of holes and electrons are equal. }
\label{fig3}
\end{figure}

The amplitude of the spontaneous signal depends on excitation intensities of the various pulses in different ways. Fig.~\ref{fig3} shows the amplitude as a function of the pulse area $A_n$ of the pulse No. $n$, defined by \mbox{$A_n=|e\vec{E}\cdot\vec{r}_{cv}/\hbar|\;\int^\infty_{-\infty}dt\, \exp[-(t/t_{Ln})^2]$}. 
A pulse with $A_n=\pi$ corresponds to complete inversion of a two--level absorber excited resonantly by pulse No. $n$. 

It is seen that in particular the half--gap contribution has to be strong enough, while the first full--gap pulse does not need to have such high intensity. While in the limit of low excitation intensity the amplitude of the signal depends linearly on the area of the first full--gap pulse, the dependence is quadratic for both the half--gap pulse and the second full--gap pulse, reflecting the lowest relevant order of the various pulses. Therefore the spontaneous intra--band response is at least of fifth order in the external light field. 

\section{Discussion}
\label{dis}
\begin{figure}
\includegraphics[width=1.5in]{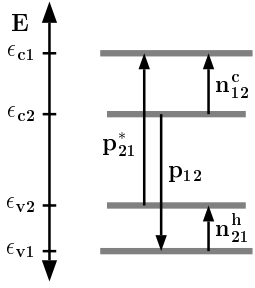}
\caption{The 4--level--model we are using in section \ref{dis} to explain how the second excitation pulse at $t=\tau$ induces spontaneous signals in the intra--band current at later times. Note that there is a dipole--moment between the two levels $v1$, $c1$ and the levels $v2$, $c2$ only. }
\label{fig4}
\end{figure}

The appearance of the spontaneous intra--band signals in systems with correlated disorder
can be easily understood on the basis of a simplified model. Let us assume that we diagonalize the conduction and valence band Hamiltonians, resulting in eigenstates $|v\nu\rangle$ and $|c\nu\rangle$ having energies $\epsilon_{v\nu}$ and $\epsilon_{c\nu}$ for the valence and conduction band, respectively. The Hamiltonian is then given in the Appendix Eqn.~(\ref{Hammi}). The optical Bloch equations describing the dynamics of the inter--band polarization $p_{\nu\nu'}=\langle a^+_{v\nu}a_{c\nu'}\rangle$ and the intra--band variables $n^h_{\nu'\nu}= \delta_{\nu\nu'}-\langle a^+_{v\nu}a_{v\nu'}\rangle$ and $n^c_{\nu\nu'}=\langle a^+_{c\nu}a_{c\nu'}\rangle$ after excitation with field $E(t)$ are given in Eqn.~(\ref{poopo}) in the Appendix. Here we are interested in the response of $n^h_{\nu'\nu}$ and $n^c_{\nu\nu'}$ to the excitation sequence. We assume the $n^h_{\nu'\nu}$ and $n^c_{\nu\nu'}$ have been created by the coherent--control pulse and are now subject to the excitation with the second full--gap pulse. This excitation is resonant in the terms proportional to the inter--band dipole matrix element $\vec\mu$ and offresonant in all terms proportional to the intra--band dipole matrix element $\vec D_{\nu\nu'}$. Thus we omit all terms proportional to $\vec D_{\nu\nu'}$ in Eqns. (\ref{pip}) and (\ref{poopop}). Note that because of the correlated disorder only inter--band dipole matrix elements $\vec\mu$ are nonzero between pairs of corresponding states, i.e. there is a strict selection rule. For tutorial reasons we consider in the following only two states in each band, i.e. $|v1\rangle$, $|v2\rangle$ and $|c1\rangle$, $|c2\rangle$, Fig.~\ref{fig4}. 
The equations of motion for $n^h_{\nu'\nu}$ and $n^c_{\nu\nu'}$ for $\nu=1\, , \, \nu'=2$ read
\begin{eqnarray}
\frac{d n^{h>}_{21}}{dt} + i\delta_v n^{h>}_{21}
&=& \frac{i}{\hbar} \vec\mu\cdot\vec E(t)
\left[p^*_{21}-p_{12}\right]\nonumber\\
\frac{d n^{c>}_{12}}{dt} - i\delta_c
n^{c>}_{12}
&=& \frac{i}{\hbar}  \vec\mu\cdot\vec E(t)
\left[p^*_{21}-p_{12}\right]
\label{pip}
\end{eqnarray}
\begin{figure}
\includegraphics[width=3.4in]{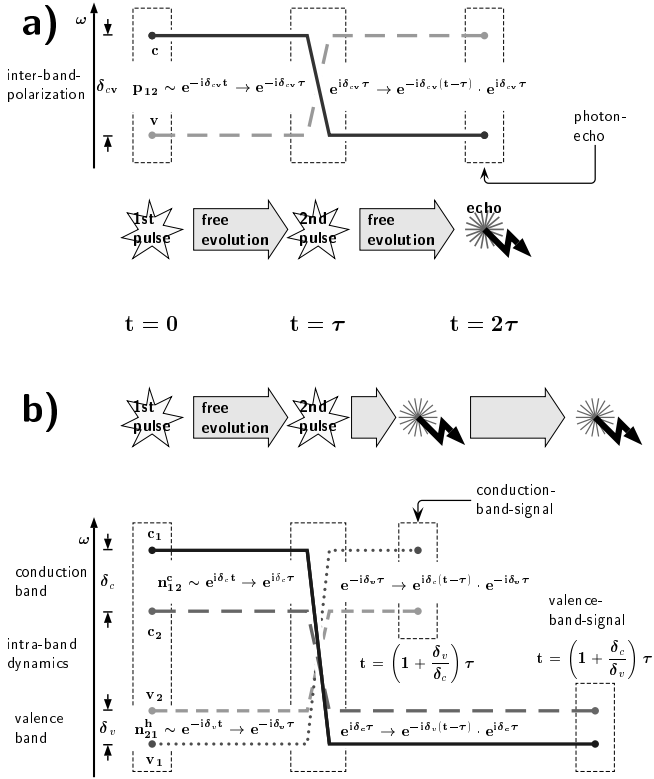}
\caption{{\bf a} The evolution of phases for the case of microscopic polarizations which cause the photon echo. 
{\bf b} Phase evolution of electron and hole intra--band coherences that give raise to the spontaneous signals of the intra--band current. Note, that the phases of electron and hole coherences are interchanged by the second excitation pulse. 
}
\label{fig5}
\end{figure}
where $\delta_v=(\epsilon_{v1}-\epsilon_{v2})/\hbar$ and $\delta_c=(\epsilon_{c1}-\epsilon_{c2})/\hbar$ with $\delta_{v, c}>0$. These equations of motion describe the intra--band dynamics after the second full--gap pulse. Just before the arrival of this second pulse the $n^{h<}_{21}$ and $n^{c<}_{12}$ have acquired phases according to $\delta_v$ and $\delta_c$, respectively, due to their free motion in the interval between the first pulse and the second pulse, see Fig.~\ref{fig5}. Just after the first excitation pulse their phases were such that (for the total ensemble of $M$ levels in each band) a macroscopic intra--band current was present. We will denote these phases as initial phases. Now our aim is to find the time when the phases of $n^{h>}_{21}$ and $n^{c>}_{12}$ again equal these initial phases. We assume for simplicity that the second pulse has the form $\delta(t-\tau)$, i.e. it arrives after a delay time $\tau$ following the first pulse (at $t=0$) and is extremely short. From the equation of motion for $p$ we find the values of this variable at time $t=\tau$ which enter the driving term on the RHS of Eqn.~(\ref{pip}):
\begin{eqnarray}
\frac{d p_{12}}{dt} + i\delta_{c2\, v1}\;p_{12}
&=& {}- \frac{i}{\hbar} \vec\mu\cdot\vec E(t)
(n^{h<}_{21} + n^{c<}_{12})\nonumber\\
\frac{d p^*_{21}}{dt} - i\delta_{c1\, v2}\;p^*_{21}
&=& \frac{i}{\hbar} \vec\mu\cdot\vec E(t)
(n^{h<}_{21} + n^{c<}_{12})
\label{poopop}
\end{eqnarray} 
where $\delta_{c2\, v1}=(\epsilon_{c2}+\epsilon_{v1})/\hbar$ and $\delta_{c1\, v2}=(\epsilon_{c1}+\epsilon_{v2})/\hbar$. If these functions, taken at time $t=\tau$, are inserted into Eqn.~(\ref{pip}), we find that the equation for $n^{h>}_{21}$ has a driving term proportional to $-n^{c<}_{12}(\tau)$ while that for $n^{c>}_{12}$ has a driving term proportional to $-n^{h<}_{21}(\tau)$. Note that due to correlated disorder the free dynamics in the upper and lower pair of states is identical up to a global scaling factor given by $\delta_v/\delta_c$. Consequently, $n^{c>}_{12}(t)$ has acquired the initial phase of $-n^{h<}_{21}(\tau)$ at time $t_c=(1+\delta_v/\delta_c)\tau$ and $n^{h>}_{21}(t)$ has acquired the initial phase of $-n^{c<}_{12}(\tau)$ at time $t_v=(1+\delta_c/\delta_v)\tau$. This is schematically shown in Fig.~\ref{fig5}b. Turning now back to the ensemble of more than two eigenstates in the bands we see that all these particular terms add up at times $t_v$ and $t_c$ to an intra--band dynamics showing the initially generated intra--band current, while all other terms interfere destructively. This explains the new intra--band phenomenon. We also learn from this consideration, i.e. the sign of the driving terms, why the spontaneous signals have the opposite sign to the first current pulse. 

In contrast, the inter--band photon echo relies on the phase conjugation of the inter--band polarizations $p_{\nu\nu'}$, as shown in Fig.~\ref{fig5}a. This inter--band--phase conjugation leads always to a restoration of the initial phases at time $t=2\tau$. If, however, one is interested in the dynamics of the intra--band quantities, we have to consider that not only the inter--band phase factors are conjugated by the second pulse, but in addition also the intra--band phases. 

\section{Conclusion}
\label{con}
\begin{figure}
\includegraphics[width=3in]{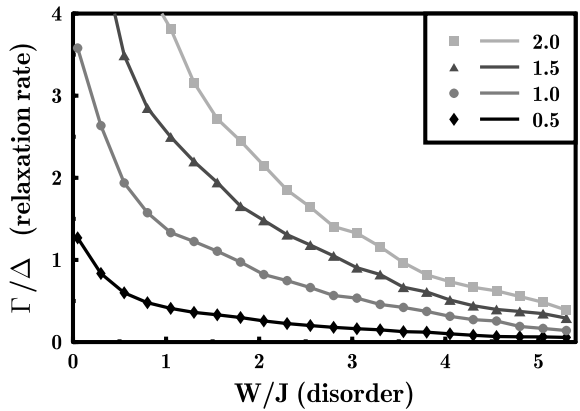}
\caption{Population relaxation rate, $\Gamma/\Delta$, as a function of
disorder, 
$W/J$, for a single--band tight--binding model with $N=20$ sites. The
initial population is a Gaussian centered around the lower band edge. $\Delta$ is the mean level 
spacing and $J$ is the nearest neighbor coupling. The curves are parameterized 
according to $U/J$, the Coulomb potential at the nearest neighbor distance.}
\label{fig6}
\end{figure}

The experimental verification or observation of the phenomena discussed in this paper should be possible in cases where dephasing interactions do not destroy the electron and hole coherences generated by the laser pulses. Interactions with the environment via phonons and for elevated excitation intensity, in particular, Coulomb--scattering provide the major causes of fast loss of phase information in these systems. Therefore it is necessary to discuss the circumstances under which these dephasing processes may still give enough time for the coherent spontaneous signals to occur. 

Coulomb--scattering in ordered or weakly disordered systems is a rapid dephasing process \cite{stephan, boris}. The question here is, how is this fact modified if disorder gets stronger. As we have seen strong disorder proves to be more effective for the realization of a current echo. It is expected that in the localized regime the Coulomb--scattering between single particle states will be strongly suppressed due to the very small overlap of these states. 

In order to have a feeling of what the interplay between strong disorder and Coulomb--scattering might be we have performed numerical simulations of a one--dimensional one--band quantum Coulomb--glass model
\cite{imili} with half--band--filling. 
After obtaining the mean--field approximation by treating the disorder exactly but the Coulomb--interaction on a Hartree--Fock level, we study the relaxation process of an initially non--equilibrium occupation probability distribution over the Hartree--Fock basis towards equilibrium caused by Coulomb--scattering. This was treated within the second order Born approximation. The initial non--equilibrium distribution meant to be `generated' by a laser pulse. 

Increasing the strength of disorder we see a very fast decrease in the relaxation rate as can be seen in Fig.~\ref{fig6}. This effect is attributed to the fact that with increasing disorder the single particle states become more localized and therefore reduce the probability of Coulomb--scattering. At the end of the relaxation process the equilibrium distribution is achieved to be a Fermi--Dirac distribution with an effective temperature that increases with the excitation energy towards the band center. 

The fact that the efficiency of Coulomb scattering can be reduced due to the presence of disorder, may make it possible to have the intra--band dephasing times sufficiently long in order to observe the spontaneous current response predicted here. Ideally suited experiments should be performed at low temperatures to reduce scattering with phonons on samples that are characterized by not to weak disorder which needs to be correlated in the valence and conduction bands. Semiconductor quantum wells with a significant amount of well width fluctuations seem to be good candidates for the experimental verification of the phenomena predicted here. 

\appendix
\section{}
The Hamiltonian for a system of $M$ states with energies $\tilde\epsilon_{\nu}^v$ being occupied in the ground state and $M$ upper unoccupied states with energies $\tilde\epsilon_{\nu}^c$ reads
\begin{equation}
H_0=\sum_{\lambda=c, v}\;\sum_{\nu=1}^{M}\tilde\epsilon_{\lambda\nu} a^{+}_{\lambda\nu}a_{\lambda\nu}. 
\label{Hammi}
\end{equation}
In the electron--hole picture $\epsilon_{\nu}^v$ and $\epsilon_{\nu}^c$ are positive quantities. In the electron picture we have $\tilde\epsilon_{\nu}^v=-\epsilon_{\nu}^v$ and $\tilde\epsilon_{\nu}^c=\epsilon_{\nu}^c$. The light--matter interaction is given by
\begin{equation}
H_L=-\vec{E}(t) \cdot \vec{d}
\label{e}
\end{equation}
with the electric field $\vec {E}(t)$ and the dipole operator
\begin{equation}
\vec{d}= \vec{\mu}\sum_{\nu=1}^{M}\left(a^{+}_{c\nu}a_{v\nu} + H. C.\right) + \sum_{\lambda=c, v}\;\sum_{\nu, \nu'=1}^{M}
\vec{D}_{\nu \nu'} a^{+}_{\lambda\nu}a_{\lambda\nu'} , 
\label{polmn}
\end{equation}
where ${\vec{D}}_{\nu \nu'}$ is the intra-band dipole matrix element between pairs of upper and lower states and $\vec{\mu}=-e\vec{r}_{cv}$. 
\begin{widetext}
\begin{eqnarray}
\frac{d p_{\nu\nu'}}{dt} + \frac{i}{\hbar}(\epsilon_{v\nu} + \epsilon_{c\nu'}) p_{\nu\nu'}
&=& {}- \frac{i}{\hbar} \vec{\mu} \cdot \vec{E}(t)\left( n^c_{\nu\nu'} + n^h_{\nu'\nu} - \delta_{\nu\nu'}\right) + \frac{i}{\hbar}\vec{E}(t) \cdot \sum_{\gamma=1}^{M} \left(\vec{D}_{\nu'\gamma}p_{\nu\gamma} - \vec{D}_{\gamma\nu}p_{\gamma\nu'}\right) , \nonumber
\\
\frac{d n^h_{\nu'\nu}}{dt} - \frac{i}{\hbar} (\epsilon_{v\nu'}-\epsilon_{v\nu}) n^h_{\nu'\nu}
&=& \frac{i}{\hbar}\vec{\mu} \cdot \vec{E}(t)\left(p^*_{\nu'\nu} - p_{\nu\nu'}\right)
+ \frac{i}{\hbar}\vec{E}(t) \cdot \sum_{\gamma=1}^{M} \left(\vec{D}_{\nu'\gamma}n^h_{\gamma\nu} - \vec{D}_{\gamma\nu}n^h_{\nu'\gamma}\right) , \nonumber
\\
\frac{d n^c_{\nu\nu'}}{dt} - \frac{i}{\hbar} (\epsilon_{c\nu}-\epsilon_{c\nu'}) n^c_{\nu\nu'}
&=& \frac{i}{\hbar}\vec{\mu} \cdot \vec{E}(t)\left(p^*_{\nu'\nu} - p_{\nu\nu'}\right) + \frac{i}{\hbar}\vec{E}(t) \cdot \sum_{\gamma=1}^{M} \left(\vec{D}_{\nu'\gamma}n^c_{\nu\gamma} - \vec{D}_{\gamma\nu}n^c_{\gamma\nu'}\right) . 
\label{poopo}
\end{eqnarray}
\end{widetext}



\end{document}